\documentclass[prb,twocolumn,showpacs,showkeys,preprintnumbers,,superscriptaddress,citeautoscript,amsmath,amssymb]{revtex4}
\usepackage{mathrsfs}
\usepackage{amssymb,amsmath}
\usepackage{graphics}
\usepackage{bm}
\usepackage{epsfig}
\usepackage{color}
\usepackage{amsfonts}

\usepackage{amscd}
\begin{document}

\def\gammav{{\mbox{\boldmath{$\gamma$}}}}
\def\sigmav{{\mbox{\boldmath{$\sigma$}}}}
\def\betav{{\mbox{\boldmath{$\beta$}}}}
\def\kappav{{\mbox{\boldmath{$\kappa$}}}}
\def\omegav{{\mbox{\boldmath{$\omega$}}}}
\def\tauv{{\mbox{\boldmath{$\tau$}}}}

\title{Filtering and analyzing mobile  qubit information via  Rashba-Dresselhaus-Aharonov-Bohm interferometers}

\author{Amnon Aharony}
\email{aaharony@bgu.ac.il} \altaffiliation{Also at Tel Aviv
University.} \affiliation{Department of Physics and the Ilse Katz
Center for Meso- and Nano-Scale Science and Technology, Ben-Gurion
University, Beer Sheva 84105, Israel}
\author{Yasuhiro Tokura}
\affiliation{ NTT Basic Research Laboratories, NTT Corporation,
Atsugi-shi, Kanagawa 243-0198, Japan}
\author{Guy Z. Cohen}
\altaffiliation{Also at Tel Aviv University.}
\affiliation{Department of Physics, Ben-Gurion University, Beer
Sheva 84105, Israel}
\author{Ora Entin-Wohlman}
\altaffiliation{Also at Tel Aviv University.}
\affiliation{Department of Physics and the Ilse Katz Center for
Meso- and Nano-Scale Science and Technology, Ben-Gurion
University, Beer Sheva 84105, Israel}
\author{Shingo Katsumoto}
\affiliation{Institute of Solid State Physics, University of
Tokyo, Kashiwa, Chiba 277-8581, Japan}

\date{\today}

\begin{abstract}
Spin-1/2 electrons are scattered through one or two diamond-like
loops, made of quantum dots connected by one-dimensional wires,
and subject to both an Aharonov-Bohm flux and (Rashba and
Dresselhaus) spin-orbit interactions. With some symmetry between
the two branches of each diamond, and with appropriate tuning of
the electric  and magnetic fields (or of the diamond shapes) this
device completely blocks electrons with one polarization, and
allows only electrons with the opposite polarization to be
transmitted. The directions of these polarizations are tunable by
these fields, and do not depend on the energy of the scattered
electrons.  For each range of fields one can tune the site and
bond energies of the device so that the transmission of the fully
polarized electrons is close to unity. Thus, these devices perform
as ideal spin filters, and these electrons can be viewed as mobile
qubits; the device writes definite quantum information on the
spinors of the outgoing electrons. The device can also read the
information written on incoming polarized electrons: the charge
transmission through the device contains full information on this
polarization. The double-diamond device can also act as a
realization of the Datta-Das spin field-effect transistor.
\end{abstract}


 \maketitle

\section{Introduction}

 Future device technology and quantum information
processing may be based on spintronics,\cite{1} where one
manipulates the electron's spin (and not only its charge). Adding
the spin degree of freedom to conventional charge-based electronic
devices  has the potential advantages of longer decoherence times
and lengths, 
increased data processing speed, lower power consumption, and
increased integration densities compared with conventional
semiconductor devices. Spins may also be used as qubits in quantum
computers.\cite{NC11} Quantum information is stored in the two
complex components of the  spinor which represents a spin-$1\over
2$ state. This information is equivalently contained in the unit
vector along which the spin is polarized. Writing and reading
information on a spin qubit is thus equivalent to polarizing this
spin along a specific direction, and later identifying this
direction. Many of the proposed experimental realizations of
qubits consider {\it static} qubits, e.g. an electron localized on
a quantum dot.\cite{loss,KBT06,tarucha} With static qubits, the
quantum information is transferred via the exchange interactions
between the qubits, rather than by the qubits themselves. Here we
consider {\it mobile} qubits:\cite{BSR00,PI04} the quantum
information is carried by polarized spin-$1\over 2$ particles
(e.g. electrons). Mobile qubits were implemented\cite{EBB00} in a
two dimensional electron gas (2DEG) using a surface acoustic wave
(SAW), that captures individual electrons along its potential
minima. Using one SAW, single electrons in parallel quantum 1D
channels can be dragged and used as synchronized inputs to a
quantum gate. Although presenting additional constraints for
coherence and synchronization, mobile qubits have many advantages
over static ones. With mobile qubits, manipulation is done by {\it
static} electric and magnetic fields rather than by expensive
high-frequency (scale of giga-Hertz) electromagnetic
pulses.\cite{HFC03}  Also, using a beam of many electrons enables
ensemble averages over the information carried by each of them,
reducing the errors.

Mesoscopic {\it spin filters} (or valves) are devices which
polarize the spins going through them along tunable directions, or
- equivalently - write quantum information on these mobile qubits.
Spin filters can also be used as {\it spin analyzers}, which read
this information by identifying the polarization directions of
incoming polarized beams. The present paper discusses such
devices. We start with a brief review of alternative approaches. A
priori, an elementary way to obtain polarized electrons is to
inject them from a ferromagnet, \cite{JKH07} after generating them
e.g. optically. \cite{OMM02} Connecting ferromagnets to
semiconductiors is inefficient, due to a large impedance mismatch
between them.\cite{ZFD04} Optical generation is difficult to
integrate with electronic devices. Another method, which also
involves ferromagnets, uses a magnetic tunnel
junction,\cite{LHS02,SM04,GBB05,LBB06} with a different
tunneling barrier height for each spin direction. 
The main difficulty is again the impedance match problem between
the ferromagnetic junction and the semiconductor at the output.
Several proposed filters use quantum dots, in which the filtering
is based on either the Coulomb blockade and the Pauli principle
\cite{RSL00,Iye,taruch} or on the Zeeman energy
splitting.\cite{FPMU03,HVB04} All the above filters usually
generate only a {\it partial} spin polarization. For writing
useful quantum information,  the outgoing electrons must be {\it
fully} polarized.

Here we follow an alternative early proposal of a spin
field-effect transistor (SFET), by Datta and Das, \cite{2} which
takes advantage of the spin-orbit interaction (SOI).  In vacuum,
the SOI has the form\cite{SN94}
\begin{align}
{\cal H}^{}_{\mathrm{SO}}=\Lambda\sigmav\cdot [{\bf p}\times
\nabla V({\bf r})],\label{eq30}
\end{align}
where  $\Lambda=\hbar/(2m^{}_0c)^2$ ($m^{}_0$ is the mass of a
free electron, $c$ is the speed of light), ${\bf p}$ is the
electron momentum, $V({\bf r})$ is the potential and the Pauli
matrices $\sigmav$ indicate the electron spin
$\mathbf{s}=\hbar\sigmav/2$. Here we concentrate on mesoscopic
structures, made of narrow gap semiconductor heterostructures, in
which electrons are confined to move in a plane (the $xy-$plane
below), forming a 2DEG. In such semiconductors, the microscopic
SOI (\ref{eq30}) modifies the band structure, and often introduces
a spin splitting of bands.\cite{winkler} The final result can
often be written as an effective SOI Hamiltonian, of the general
form
${\cal H}^{}_{\mathrm{SO}}=(\hbar /m)(\kappav^{}_{\mathrm
SO}\cdot\sigmav)$,  
where $\kappav^{}_{\mathrm SO}$ is a linear combination of the
electron momentum components $p^{}_x$ and $p^{}_y$ and $m$ is the
effective mass, which is usually much smaller than $m^{}_0$. The
related energy scale can be larger than that of Eq. (\ref{eq30})
by as much as six orders of magnitude.

The literature has emphasized two special cases of the effective
SOI. A confining potential well which is asymmetric under space
inversion generates the Rashba SOI. \cite{3} For an electric field
${\bf E}=-\nabla V$ in the $z-$direction, this SOI is similar to
Eq. (\ref{eq30}):
\begin{align}
{\cal H}^{}_{\mathrm{R}}=\frac{\hbar
k^{}_{R}}{m}(p^{}_y\sigma^{}_x-p^{}_x\sigma^{}_y).\label{Rashba}
\end{align}
The coefficient $k^{}_R$ typically depends on ${\bf E}$, as indeed
confirmed experimentally.\cite{koga,rsoi1,rsoi2,koga06} When the
bulk crystal unit cell lacks inversion symmetry, one also has the
Dresselhaus SOI,\cite{D55} which is usually cubic in the momentum.
For a 2DEG this SOI is given by
\begin{align}
{\cal H}^{}_{\mathrm{D}}=\frac{\hbar
k^{}_{D}}{m}(p^{}_x\sigma^{}_x-p^{}_y\sigma^{}_y),\label{Dress}
\end{align}
where $k^{}_D$ usually depends on the crystal structure and only
weakly (if at all) on the external field. From the strictly
theoretical point of view, however, there is not much difference
between the linear Dresselhaus interaction and the Rashba term, as
they are connected by a unitary transformation.\cite{D1} A similar
transformation can also switch the sign of the second term in
(\ref{Dress}). For the purposes of the present paper we shall keep
the generic separation between ${\cal H}^{}_{\mathrm R}$ and
${\cal H}^{}_{\mathrm D}$.

When a spin moves a distance $L$ in the direction of the unit
vector $\hat{\bf g}$ then its spinor $|\chi\rangle$ transforms
into $|\chi\rangle \rightarrow U|\chi\rangle$, with the unitary
spin rotation matrix\cite{oreg}
\begin{align}
U=e^{i{\bf K}\cdot\sigmav},\ \ {\bf
K}=\alpha^{}_R(g^{}_y,-g^{}_x,0)+\alpha^{}_D(g^{}_x,-g^{}_y,0),
\label{oreg1}
\end{align}
with $\alpha^{}_{R,D}=k^{}_{R,D}L$. The Datta-Das  SFET used this
effect to rotate the spins of electrons which move in a
quasi-one-dimensional semiconductor wire, connected to two
ferromagnets. Experimental realizations of this device are still
awaiting the solution of the impedance matching problem. From now
on we discuss filters which avoid ferromagnets. In a 2DEG with
SOI, an interface between two regions with different SOI's causes
a splitting of each beam  into two polarized beams with different
velocities. This was the basis for
  the refraction/reflection filter.\cite{KSF04,SKF05,LYS10}
  Another SOI based filter uses mesoscopic T junctions, which
split the unpolarized electron beam into two polarized
ones.\cite{YOO05,YOK05,YDKO06,YE10} These filters are advantageous
since they produce two polarized beams, thus using all the
electrons in the original beam, and since they do not use magnetic
fields. However, the outgoing polarization depends on the
electrons' energy.  In most of this paper we calculate the
transmission of electrons, moving from a left lead to a right lead
via a scattering device, the `filter'. However, at the end we also
mention the conductance between two unpolarized reservoirs which
are connected to these leads, and then one may need to average the
polarization of electrons with different energies, e.g. at finite
temperature and/or finite bias voltage. It is thus advantageous to
have {\it energy-independent} polarizations.

The filters discussed below take advantage of the {\it
interference of electronic waves in quantum networks} which
contain closed loops. The phases of these waves can include the
Aharonov-Bohm (AB) phase $\phi$,\cite{AB} which results from a
magnetic flux $\Phi$ penetrating each loop. When an electron goes
around a loop its wave functions gains an AB phase $\phi\equiv 2
\pi\Phi/\Phi^{}_0$, and $\Phi_0=hc/e$ is the unit flux ($e$ is the
electron charge). This phase is a special example of the Berry
phase.\cite{berry} Another example involves the Aharonov-Casher
effect,\cite{AC} which is related to the spin degree of freedom.
When an electron goes around a loop along which it is subject to
the SOI, its spinor rotates by the transformation
\begin{align}
u=\exp[i\omegav\cdot\sigmav]\equiv\cos\omega+i\sin\omega\hat{\bf
m}, \label{uu} \end{align} where $\hat{\bf
m}\equiv\omegav/\omega$, introducing an additional SOI-related
phase $\omega=|\omegav|$. The matrix $u$ is a product of matrices
of the kind given in Eq. (\ref{oreg1}), each coming from the local
SOI on a segment of the loop. As indicated by Eq. (\ref{oreg1}),
this matrix depends on the geometric details  of the bonds around
the loop (unlike $\phi$, which only depends on the area of the
loop). Indeed, many papers proposed a single circular loop
interferometer which would be sensitive to this phase and/or to
its competition with the AB
phase.\cite{oreg,ora,aronov,nitta,molnar,FR04,citro,RCFP,VKN,hatano}
The loop is connected to two leads, and the destructive
interference of the waves in the two paths can sometimes block
electrons with one polarization, and fully transmit electrons with
the opposite polarization. Some papers also suggested to connect
the loop to three leads, as in a Stern-Gerlach
experiment.\cite{FKBP08,CZ08} However, the calculated criteria for
filtering in these papers were usually energy-dependent, and there
was no systematic discussion of these criteria and of the
polarization of the transmitted spins. An alternative geometry
replaces the circular loop by a diamond-shaped square, with a SOI
on its four edges (which determine $\omegav$) and with a
penetrating AB flux, see Fig. 1.\cite{hatano,CC08} Indeed, these
papers find criteria for full spin filtering, but restrict their
discussion to isolated values of the AB flux and the SOI. Below we
generalize these pioneering results in many directions.

Interference becomes simpler in the Mach-Zhender mesoscopic
interferometer, which imitates the two-slit experiment.
\cite{ID03,k1,z1,ZS07,LMBB10} Reference
~\onlinecite{LMBB10} found
an energy-independent criterion (which relates the SOI strength
and the AB flux) for full spin polarization and an
energy-independent polarization direction, similar to those
discussed below. Since the Mach-Zhender interferometer requires
two beam splitters, which may not be easy to realize, we consider
mainly simple interferometers, based on one or two loops. Networks
of rings have also been considered, with SOI and (sometimes) with
an AB flux.\cite{MVP05,bercioux,koga06,AETK08,AETK09,kalman} In
particular, an infinite network of diamond-shaped
rings\cite{AETK08,AETK09} was found to give a wide range of
electric and magnetic fields with full polarization at the output.
Since infinite networks are difficult to realize, we discuss here
only the cases of one and two diamonds.

\begin{figure}[h]
\includegraphics[width=7.5 cm]{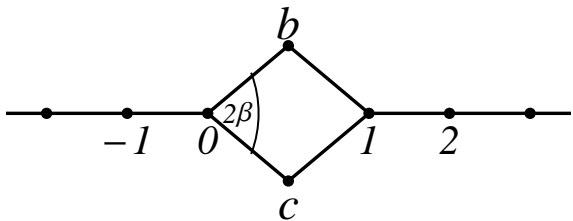}
 \caption{The single diamond and the
leads. The diamond is penetrated by a magnetic flux $\Phi$, and
the bonds around it are subject to SOI's.}\label{1}
\end{figure}

\begin{figure}[h]
\vspace{.3cm}
\includegraphics[width=8 cm]{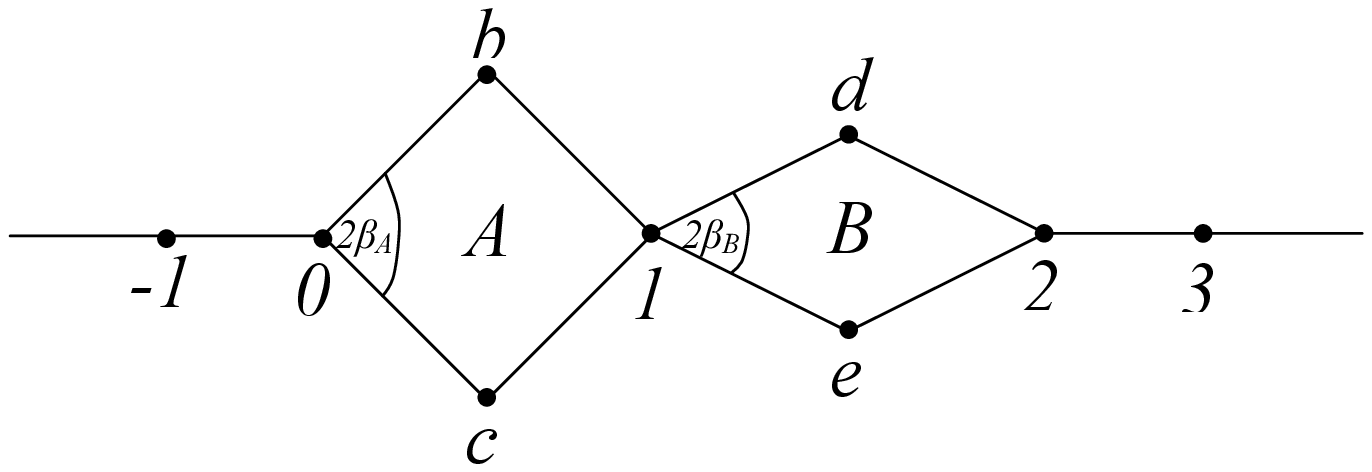}\\
\includegraphics[width=8 cm]{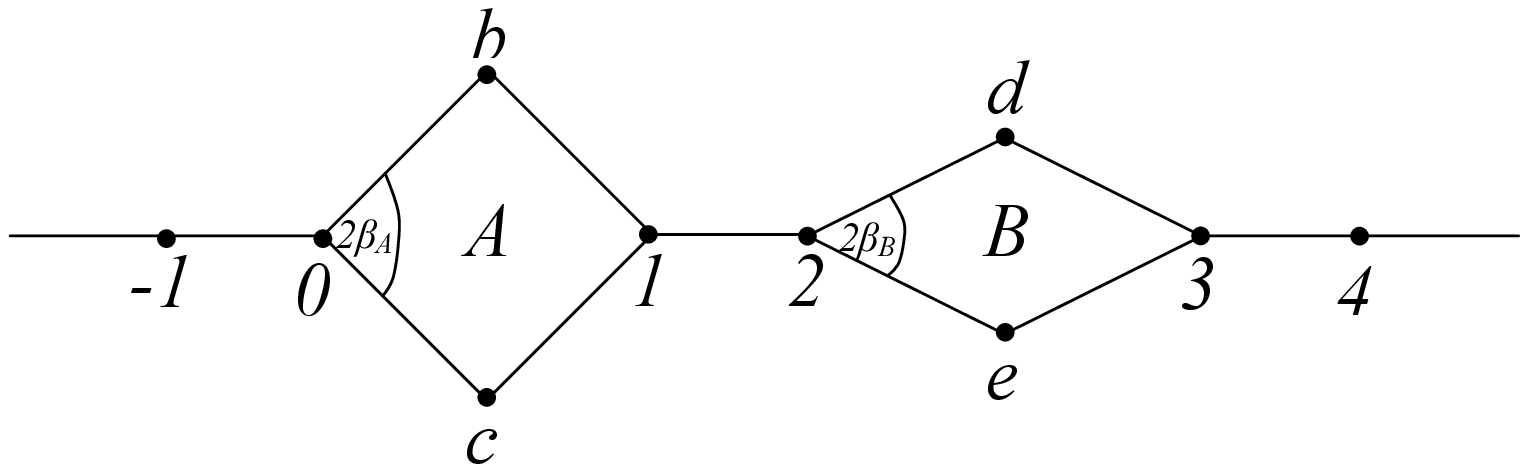}
 \caption{Schematic diagrams of the two-diamond filters. }\label{2f}
\end{figure}

\begin{figure}[h]
\includegraphics[width=5.5 cm]{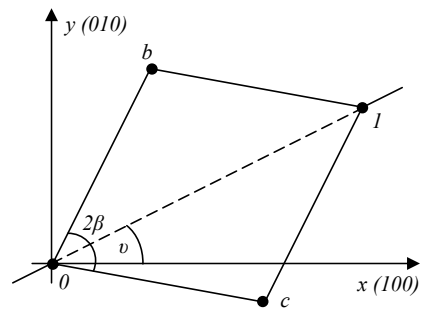}
 \caption{Schematic diagram of the rotated diamond, in the presence of the Dresselhaus SOI. The $x$ and $y$ axes
 are set at the crystal axes of the material. }\label{5}
\end{figure}

In this paper we avoid some of the problems listed above. Using
scattering theory, we calculate the spin-dependent transmission
through a single and a double diamond-like loops (Figs. \ref{1}
and \ref{2f}). We allow for general opening angles of the diamond
rhombi, $\{2\beta^{}_i\}$, which affect both the SOI, via the
lengths and orientations of the bonds, and the AB flux, via the
diamond area. We also include both the Rashba and the Dresselhaus
SOI, in addition to the AB flux. The Dresselhaus SOI depends on
the relative rotation of the crystal axes and the diamond bonds,
see Fig. \ref{5}. For a fixed value of $k^{}_D$ we find explicit
and relatively simple relations between $\phi$, $k^{}_R$ and the
$\beta$'s, at which the transmitted electrons are fully polarized
in tunable directions which we calculate. These relations and the
spin polarizations do not depend on the energy of the electrons.
The transmission coefficient of these polarized electrons can be
tuned to be very close to unity. The transmission of electrons
with other polarizations is smaller, and can be used for `reading'
their polarization.

Section II discusses the single-diamond case, Fig. 1. We first
present a general calculation of the the transmission through the
diamond, valid for any internal structure of the two
one-dimensional paths (Subsec. IIA), find the general conditions
for full filtering and for using the filter as an analyzer (IIB),
and then find specific criteria for the Rashba-only SOI (IIC) and
for both Rashba and Dresselhaus SOI's (IID). Section III discusses
two diamonds in series, as in Fig. \ref{2f}, (IIIA) and gives
explicit expressions for special cases which can yield full
filtering with a transmission of unity for practically all
energies (IIIB). The results are discussed and summarized in Sec.
IV.

 \section{A single diamond}

 \subsection{Transmission of an arbitrary diamond}

We start with the scattering of an electron from a diamond with
arbitrary SOI and AB flux. Consider an electron with spin $1\over
2$, moving on a general network of sites. The tight-binding
Schr\"odinger equation for the spinor
 $|\psi(u)\rangle$ at site $u$
is written as \begin{align}
(\epsilon-\epsilon_u)|\psi(u)\rangle=-\sum_v
\widetilde{U}^{}_{uv}|\psi(v)\rangle,
\end{align}
where $v$ runs over the nearest neighbors of $u$, while
$\widetilde{U}^{}_{uv}\equiv J^{}_{uv}U^{}_{uv}$, $J^{}_{uv}$ is a
 hopping energy and $U^{}_{uv}$ is a $2\times 2$ unitary
matrix. For the diamond in Fig. 1, the matrices $U^{}_{uv}$ differ
from the $2\times 2$ unit matrix ${\bf 1}$ only for the four bonds
forming the diamond. At this stage we do not specify the details
of these four matrices, which contain the AB phase and the SOI
rotation, or of the corresponding four coefficients $J^{}_{uv}$.

Except for the above four bonds, the nearest neighbor hopping
energy along the leads is $j$, with no spin-orbit interaction, and
the site energies $\epsilon^{}_u$ on the leads are zero. With a
lattice constant $a$, the states on the leads are combinations of
$e^{\pm inka}$, multiplying $n-$independent spinors, and the
corresponding energy is $\epsilon=-2 j \cos(ka)$.
 The Schr\"odinger equations for the spinors at the
corners of the diamond are
\begin{align}
&(\epsilon-\epsilon^{}_0)|\psi(0)\rangle=-\bigl
(\widetilde{U}^{}_{0b}|\psi(b)\rangle+\widetilde{U}^{}_{0c}|\psi(c)\rangle\bigr
)-j|\psi(-1)\rangle,\nonumber\\
&(\epsilon-\epsilon^{}_1)|\psi(1)\rangle=-\bigl
(\widetilde{U}^{\dagger}_{b1}|\psi(b)\rangle+\widetilde{U}^{\dagger}_{c1}|\psi(c)\rangle\bigr
)-j|\psi(2)\rangle,\nonumber\\
 &(\epsilon-\epsilon^{}_b)|\psi(b)\rangle=-\bigl
(\widetilde{U}^\dagger_{0b}|\psi(0)\rangle+\widetilde{U}^{}_{b1}|\psi(1)\rangle\bigr ),\nonumber\\
&(\epsilon-\epsilon^{}_c)|\psi(c)\rangle=-\bigl
(\widetilde{U}^\dagger_{0c}|\psi(0)\rangle+\widetilde{U}^{}_{c1}|\psi(1)\rangle\bigr
).
\end{align}
Substituting the last two equations into the first two, one has
\begin{align}
&(\epsilon-y^{}_0)|\psi(0)\rangle={\bf
W}|\psi(1)\rangle-j|\psi(-1)\rangle,\nonumber\\
&(\epsilon-y^{}_1)|\psi(1)\rangle={\bf
W}^\dagger|\psi(0)\rangle-j|\psi(2)\rangle,\label{66}
\end{align}
where \begin{align}
 y^{}_u\equiv
\epsilon^{}_u+\gamma^{}_{ubu}+\gamma^{}_{ucu},
 \ \  \gamma^{}_{uvw}\equiv
J^{}_{uv}J^{}_{vw}/(\epsilon-\epsilon_v),\label{defz} \end{align}
\begin{align}
&{\bf
W}\equiv\gamma^{}_{0b1}U^{}_{0b}U^{}_{b1}+\gamma^{}_{0c1}U^{}_{0c}U^{}_{c1}.\label{WW}
\end{align}
Generally, ${\bf W}$ is not a unitary matrix (unlike the $U$'s).
Like any $2\times 2$ matrix, ${\bf W}$ can always be written as
\begin{align}
{\bf W}=d+{\bf b}\cdot\sigmav,\label{WW1}\end{align} where $d$ and
${\bf b}$ are a complex number and a complex three-component
vector, which are determined by the details of the hopping
matrices $U^{}_{uv}$.

A wave coming from the left has the form
\begin{eqnarray}
|\psi(n)\rangle&=&e^{ikna}|\chi^{}_{in}\rangle+r e^{-ikna}|\chi^{}_r\rangle,\ \ n\le 0,\nonumber\\
|\psi(n)\rangle&=&t e^{ik(n-1)a}|\chi^{}_t\rangle,\ \ n\ge 1,
\label{psi}\end{eqnarray}
where $|\chi^{}_{in}\rangle$,  $|\chi^{}_r\rangle$ and
$|\chi^{}_t\rangle$ are the incoming, reflected and transmitted
normalized spinors, respectively  (with the corresponding
reflection and transmission complex amplitudes $r$ and $t$).
Substituting Eqs. (\ref{psi}) into Eqs. (\ref{66}) one finds
\begin{align}
t|\chi^{}_t\rangle={\cal T}|\chi^{}_{in}\rangle,\ \ \
r|\chi^{}_r\rangle={\cal R}|\chi^{}_{in}\rangle, \end{align}
 with
the $2\times 2$ transmission and reflection amplitude matrices
\begin{align}
{\cal T}&=2ij\sin(ka){\bf W}^\dagger\bigl (Y{\bf 1}-{\bf
WW}^\dagger\bigr )^{-1},\\
%
{\cal R}&=-{\bf 1}-2ij\sin(ka)X^{}_1(Y{\bf 1}-{\bf W}{\bf
W}^\dagger)^{-1}.\label{eq02} \end{align} Here,
\begin{align}
Y=X^{}_0X^{}_1,\ \ X^{}_u=y^{}_u +je^{-ika}.\label{YX}
\end{align}

Both ${\cal T}$ and ${\cal R}$ involve the hermitian matrix
\begin{align}
{\bf WW}^\dagger=A+{\bf B}\cdot \sigmav, \label{eq25} \end{align}
where [by Eq. (\ref{WW1})]
\begin{align}
&A=|d|^2+{\bf b}\cdot{\bf b}^\ast,\nonumber\\
&{\bf B}=2{\rm Re}[d^\ast{\bf b}]+2[{\rm Re}({\bf b})\times{\rm
Im}({\bf b})]\equiv |{\bf B}|\hat{\bf n}. \label{WWA}\end{align}
Defining the eigenstate of the spin component along a general unit
vector $\hat{\bf n}$ via
$\hat{\bf n}\cdot \sigmav |\hat{\bf n}\rangle= |\hat{\bf
n}\rangle$,  
the eigenvectors of ${\bf WW}^\dagger$ are identified as
$|\pm\hat{\bf n}\rangle$,
 \begin{align}{\bf WW}^\dagger|\pm\hat{\bf
n}\rangle=\lambda^{}_\pm|\pm\hat{\bf n}\rangle,\ \ \
\lambda^{}_\pm=A\pm|{\bf B}|.\end{align}

 Equation (\ref{WW}) presents an example of the general two-path loop, for which one can write
\begin{align}
{\bf W}=\gamma^{}_bU^{}_b+\gamma^{}_cU^{}_c,\label{genW}
\end{align}
with real coefficients $\gamma^{}_b$ and $\gamma^{}_c$ and
 with unitary matrices $U^{}_b$ and $U^{}_c$
corresponding to the two paths.   The same form (\ref{genW}) is
found when each path contains a chain of many bonds in
series.\cite{AETK09} This form yields
\begin{align}
{\bf W}{\bf
W}^\dagger=\gamma^2_b+\gamma^2_c+\gamma^{}_b\gamma^{}_c(u+u^\dagger),\label{WWu}
\end{align}
where $u\equiv U^{}_{b}U^{\dagger}_{c}$ is the unitary matrix
representing hopping from 0 back to 0 around the
loop.\cite{hatano} As discussed in the introduction, this matrix
has the form $u=e^{-i\phi+i\omegav\cdot\sigmav}$, see Eq.
(\ref{uu}) and preceding discussion.  Thus,
$u+u^\dagger=2(\cos\omega\cos\phi+\sin\omega\sin\phi\hat{\bf
m}\cdot\sigmav)$, and one identifies $\hat{\bf m}=\hat{\bf n}$ and
\begin{align}
&A=\gamma^2_b+\gamma^2_c+2\gamma^{}_b\gamma^{}_c\cos\omega\cos\phi,\nonumber\\
&{\bf B}=2\gamma^{}_b\gamma^{}_c\sin\omega\sin\phi\hat{\bf
n}.\label{AB}\end{align} The eigenvalues $\lambda^{}_\pm$ now
become
\begin{align}
\lambda^{}_\pm=A\pm|{\bf
B}|=\gamma^2_b+\gamma^2_c+2\gamma^{}_b\gamma^{}_c\cos(\phi\pm\omega).\label{newlam}
\end{align}
The corresponding eigenstates, $|\pm\hat{\bf n}\rangle$, represent
electrons which are fully polarized along $\pm\hat{\bf n}=\pm{\bf
B}/|{\bf B}|$. The direction of $\hat{\bf n}$ depends on the sign
of $\sin\omega\sin\phi$, namely on the directions of the magnetic
field (determining the sign of $\phi$) and of the electric field
(determining the sign of $\omega$ in the Rashba case). Switching
the sign of $\phi$ or of $\omega$ switches the direction of the
polarized spins associated with the two eigenvalues.

Equation (\ref{eq02}) implies that an incoming spinor $|\pm{\bf
n}\rangle$ will generate an outgoing spinor
\begin{align}
t|\chi^{out}_\pm\rangle={\cal T}|\pm{\bf
n}\rangle=\frac{2ij\sin(ka)}{Y-\lambda^{}_\pm}{\bf
W}^\dagger|\pm{\bf n}\rangle.
\end{align}
Since the scalar product of ${\bf W}^\dagger|\pm{\bf n}\rangle$
with itself equals $\lambda^{}_\pm$, it follows that
\begin{align}
|\chi^{out}_\pm\rangle={\bf W}^\dagger|\pm{\bf
n}\rangle/\sqrt{\lambda^{}_\pm}\label{chout},
\end{align} and that the corresponding transmission amplitude is
\begin{align}
t^{}_\pm=\frac{2ij\sin(ka)}{Y-\lambda^{}_\pm}\sqrt{\lambda^{}_\pm},
\label{trans}
\end{align}
which is an eigenvalue of ${\cal T}$.

Equation ({\ref{chout}) also implies the relation
 ${\bf W}^\dagger{\bf
W}|\chi^{out}_\pm\rangle=\lambda^{}_\pm|\chi^{out}_\pm\rangle$,
showing that $|\chi^{out}_\pm\rangle$ is an eigenstate of
\begin{align}
{\bf W}^\dagger{\bf W}=A+{\bf B}'\cdot\sigmav,\label{eq25a}
\end{align}
 where
\begin{align}
{\bf B}'=2{\rm Re}[d^\ast{\bf b}]-2[{\rm Re}({\bf b})\times{\rm
Im}({\bf b})]\equiv |{\bf B}|\hat{\bf n}'.
\label{Bprime}\end{align} Therefore, $|\chi^{out}_\pm\rangle$
corresponds to a spin direction $\hat{\bf n}'$, which differs from
$\hat{\bf n}$ in that the component along $[{\rm Re}({\bf
b})\times{\rm Im}({\bf b})]$ is reversed.  One can thus identify
$|\pm\hat{\bf n}'\rangle$ as the left eigenstates of ${\bf
W}^\dagger$, namely
\begin{align}
{\bf W}^\dagger\equiv\sqrt{\lambda^{}_-}|-{\bf n}'\rangle\langle
-{\bf n}|+\sqrt{\lambda^{}_+}|\hat{\bf n}'\rangle\langle \hat{\bf
n}|.
\end{align}
Similarly, \begin{align} {\cal T}\equiv t^{}_-|-{\bf
n}'\rangle\langle -{\bf n}|+t^{}_+|\hat{\bf n}'\rangle\langle
\hat{\bf n}|.\label{eigenT}
\end{align}

Scattering from the right lead to the left leat is obtained by
replacing ${\bf W}^\dagger$ by ${\bf W}$. It follows that an
electron polarized along $\hat{\bf n}'$ coming from the right hand
side (RHS) exits to the left hand side (LHS) polarized along
$\hat{\bf n}$. It is now straightforward to find the transmission
and reflection matrices ${\cal T}'$ and ${\cal R}'$ for this
reversed scattering: all one needs to do is interchange ${\bf W}$
with ${\bf W}^\dagger$ and $X^{}_0$ with $X^{}_1$. Note that
generally ${\cal T}'\ne{\cal T}$; these matrices are related to
each other via the self-duality of the scattering
matrix.\cite{beenaker} It is then straightforward to  check
unitarity, e.g. ${\cal T}^\dagger{\cal T}+{\cal R}'^\dagger{\cal
R}'={\bf 1}$.

\subsection{Ideal filter and reader}

Many earlier papers considered the polarization of the moving
electrons along a particular fixed direction, e.g. along the
$z-$axis. Following e.g. Ref.
~\onlinecite{hatano}, we find it much
better to consider the polarization along a {\it tilted}
direction, associated with the eigenstates of the matrix ${\bf
W}{\bf W}^\dagger$ [or, equivalently, of the matrix $u+u^\dagger$,
Eq. (\ref{WWu})]. A general incoming spinor $|\chi^{}_{in}\rangle$
can be expanded in terms of these basis eigenvectors,
\begin{align}
|\chi^{}_{in}\rangle=c^{}_+|\hat{\bf n}\rangle+c^{}_-|-\hat{\bf
n}\rangle,\label{chipm}
\end{align} with $c^{}_\pm=\langle \pm\hat{\bf n}|\chi^{}_{in}\rangle$, and then the outgoing spinor becomes
\begin{align}
t|\chi^{}_t\rangle=c^{}_+ t^{}_+|\hat{\bf n}'\rangle+c^{}_-
t^{}_-|-\hat{\bf n}'\rangle.\label{mix}
\end{align}
The total charge transmission is therefore
$T=|c^{}_+|^2T^{}_++|c^{}_-|^2T^{}_-$, with $T^{}_\pm\equiv
|t^{}_\pm|^2$ being the eigenvalues of ${\cal T}{\cal T}^\dagger$.
Given Eqs. (\ref{newlam}) and (\ref{trans}), $T^{}_\pm$ is a
function of $\phi\pm\omega$.\cite{ora} Note that $t^{}_\pm$ and
$T^{}_\pm$ are the eigenvalues of ${\cal T}$ and of ${\cal T}{\cal
T}^\dagger$, respectively.

 The single diamond
described above can serve as  a perfect filter if one of the
eigenvalues $\lambda^{}_\pm$, say $\lambda^{}_-$, vanishes. In the
following sections we show that there exist physical parameters
for which this can be achieved - independently of the electron
energy $\epsilon$. Indeed, if $\lambda^{}_-=0$ then one also has
$t^{}_-=0$, and Eq. (\ref{mix}) reduces to
$t|\chi^{}_t\rangle=c^{}_+t^{}_+|\hat{\bf n}'\rangle$.
All outgoing electrons are then polarized along $\hat{\bf n}'$,
and the total transmission strength, i.e. the  fraction of the
incoming current which exits on the RHS, is given by $T=
T^{}_+|c^{}_+|^2$.

From Eq. (\ref{newlam}) it follows that $\lambda^{}_\pm\ge 0$, and
that the equality $\lambda^{}_-=0$ can occur {\it only} if
\begin{align}
\gamma^{}_b=\gamma^{}_c\equiv\gamma\ \  {\rm and}\ \
\cos(\phi-\omega)=-1.\label{condi}\end{align}
 The first relation implies a symmetry between the two paths. For the
 specific diamond geometry of Fig. 1, one has
$\gamma^{}_v\equiv\gamma^{}_{0v1}$ [see Eqs. (\ref{defz}) and
(\ref{WW})]. If one imposes the symmetric relation
$J^{}_{0b}J^{}_{b1}=J^{}_{0c}J^{}_{c1}$, then this condition
requires $\epsilon^{}_b=\epsilon^{}_c$. Both the $J$'s and the
$\epsilon^{}_u$'s can be tuned via appropriate gate voltages, so
that the equality $\gamma^{}_b=\gamma^{}_c$ can be achieved. The
second condition in Eq. (\ref{condi}), namely $\omega=\phi+\pi$,
yields a relation between the AB flux and the SOI strength
(represented by $\omega$). Note that $\phi$ and $\omega$ depend
only on the unitary matrices $U^{}_{uv}$, and {\it not} on the
energy $\epsilon$ nor on the site energies $\epsilon^{}_u$. Also,
the vectors $\hat{\bf n}$ and $\hat{\bf n}'$ depend only on the
parameters in these matrices. Thus, for fixed diamond parameters
which obey the above conditions the direction of the outgoing
electrons' polarization is independent of the energy, and remains
the same even after summation over energies due to finite
temperature or
bias voltage (see below). 

 Substituting $\omega=\phi+\pi$
and $\gamma^{}_b=\gamma^{}_c=\gamma$ in Eq. (\ref{newlam})  yields
\begin{align}
\lambda^{}_+=4\gamma^2\sin^2\phi.\label{lam+}\end{align} This
result for $\lambda^{}_+(\epsilon)$ is {\it universal}, in the
sense that it depends on the parameters of the diamond only
through the AB flux, and {\it not} on the angle of opening $\beta$
(Fig. 1) nor on the SOI strengths $k^{}_R$ and $k^{}_D$. Of
course, these latter parameters still need to be adjusted by Eq.
(\ref{condi}) to achieve full polarization. Having satisfied Eq.
(\ref{condi}), the transmission becomes [Eq. (\ref{trans})]
\begin{align}
T^{}_{+}(\epsilon)=\frac{4j^2\sin^2(ka)\lambda^{}_+}{P+Q\cos(ka)+R\cos(2ka)},\label{Tp}
\end{align}
where
\begin{align}
P&=(y^{}_0y^{}_1-\lambda^{}_+)^2+(y^{}_0+y^{}_1)^2j^2+j^4,\nonumber\\
Q&=2j(y^{}_0y^{}_1-\lambda^{}_++j^2)(y^{}_0+y^{}_1),\nonumber\\
R&=2j^2(y^{}_0y^{}_1-\lambda^{}_+),
\end{align}
and  one can read the transmission from graphs of $T^{}_+$ as
function of $\phi$ and $\epsilon$. When $\sin\phi=0$ both
$\lambda^{}_+$ and $\lambda^{}_-$ vanish when also $\omega=\pi$,
and {\it all} the electrons are fully reflected. Therefore, one
needs $\sin\phi\ne 0$, i.e. a non-zero magnetic field. However, as
we show below, one can achieve good filtering even for small
magnetic fields.

Although all these results are specific for the tight-binding
model, one would like to apply them for general leads, with
general dispersion relations. For this purpose, it is customary to
calculate the tight-binding transmission for energies near the
center of the band, $\epsilon=0$ or $ka=\pi/2$, where the density
of states 
is flat. 
Equation (\ref{lam+}) shows
that (for $\phi\ne 0$) $\lambda^{}_+$ diverges as
$\gamma^2\propto(\epsilon-\epsilon^{}_b)^{-2}$ at the resonant
energy $\epsilon=\epsilon^{}_b$. This yields a Fano-like zero of
$T^{}_{+}$ for this energy. Since we prefer to have a weak energy
dependence around $\epsilon=0$, it is  preferable to have a
non-zero (and large) site energy $\epsilon^{}_b$.

From now on we set $J^{}_{uv}\equiv J$, so that also
$\gamma^{}_{0v0}=\gamma^{}_{1v1}=\gamma$ for $v=b,c$ [see Eq.
(\ref{defz}) and Fig. 1]. At the band center ($\epsilon=0$ or
$ka=\pi/2$), we also have
$\gamma\rightarrow\gamma^{}_0=-J^2/\epsilon^{}_b$, and the
denominator in Eq. (\ref{Tp}) becomes
$P-R=[(\epsilon^{}_0+2\gamma^{}_0)(\epsilon^{}_1+2\gamma^{}_0)-\lambda^{}_+-j^2]^2
+j^2(\epsilon^{}_0+\epsilon^{}_1+4\gamma^{}_0)^2$, which is
minimal at $\epsilon^{}_0=\epsilon^{}_1=-2\gamma^{}_0\equiv
2J^2/\epsilon^{}_b$. In this case one has
$T^{}_+=4j^2\lambda^{}_+/(\lambda^{}_++j^2)^2$, and this has its
maximal value of 1 at $\lambda^{}_+=j^2$. For a specific filter
one would usually decide over what range of flux $\phi$ one would
like to work. Fixing $j$ and $J^{}_{uv}=J$ and denoting the middle
of that range of $\phi$ by $\phi^{}_0$, one has $T^{}_+=1$ at
$\phi=\phi^{}_0$ if one tunes the parameters so that
$\gamma=\gamma^{}_0=j/(2\sin\phi^{}_0)$, and
$\epsilon^{}_b=-J^2/\gamma^{}_0,~\epsilon^{}_0=\epsilon^{}_1=-2\gamma^{}_0$.
For these choices, one ends up with
\begin{align}
T^{}_+(\epsilon=0)=4\sin^2\phi\sin^2\phi^{}_0/(\sin^2\phi+\sin^2\phi^{}_0)^2,\label{TTphi}
\end{align}
depending only on $\phi^{}_0$, as shown on the LHS of Fig.
\ref{Tpp}.   $T^{}_+(0)$ has a reasonably flat maximum at
$\phi=\phi^{}_0$ (and a width which increases with $\phi^{}_0$).
The other panel in Fig. \ref{Tpp} shows $T^{}_+$ versus $ka$ for
the flux fixed at $\phi=\phi^{}_0$, for the site energies chosen
above and for $J=4j$. As expected, $T^{}_+$ is practically
energy-independent and equal to unity in a range around the band
center $ka=\pi/2$. The width of these plateaus increases with
increasing $J$, when $|\epsilon^{}_b|=J^2/|\gamma^{}_0|\gg
|\epsilon|$.

\begin{figure}[h]
\includegraphics[width=4.1 cm]{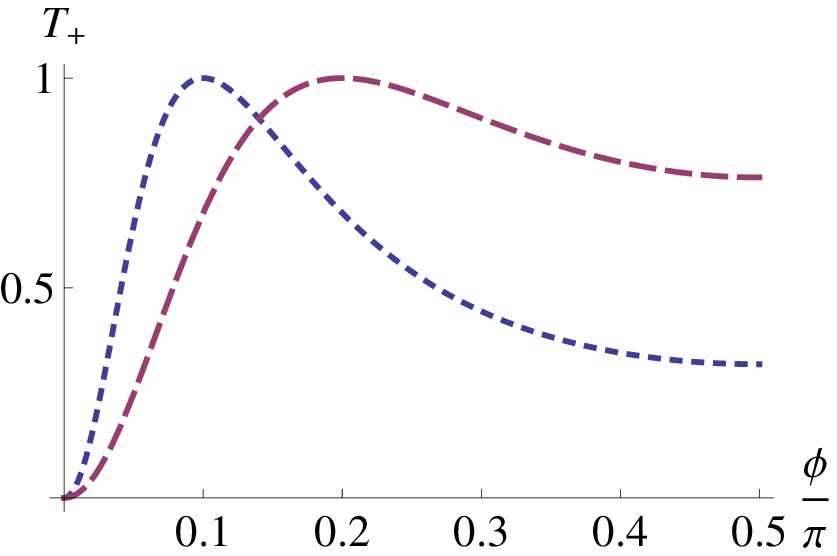}\ \ \
\includegraphics[width=4.1 cm]{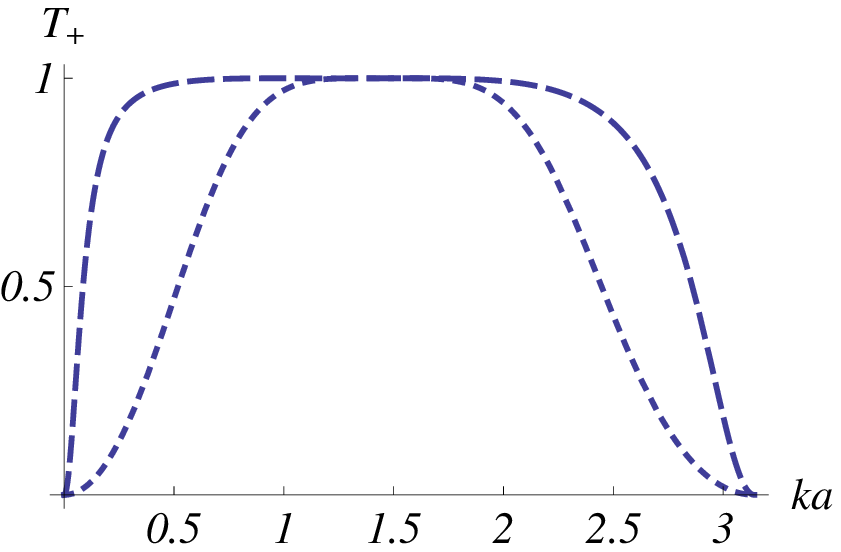}
 \caption{The transmission of the polarized electrons, $T^{}_+(\epsilon)$. LHS:  in the band
 center ($\epsilon=0$)
  versus the AB flux $\phi$
 (in units of $\pi$).  RHS:  versus $ka$ [the electron energy is $\epsilon=-2j\cos(ka)$], for
 hopping strengths around the diamond $J^{}_{uv}=J=4j$ and site energies $\epsilon^{}_0=\epsilon^{}_1=-j/\sin\phi^{}_0,
 ~\epsilon^{}_b=\epsilon^{}_c=-2J^2\sin\phi^{}_0/j$. These site energies are chosen so that $T^{}_+$ is maximal at
 $\phi=\phi^{}_0$. Small (large) dashes
 correspond to maxima of $T^{}_+(0)$ at $\phi^{}_0=.1 \pi~(.2 \pi)$.}\label{Tpp}
\end{figure}

If the incoming electrons are polarized along a direction
$\hat{\bf n}^{}_0$, namely $|\chi^{}_{in}\rangle\equiv |\hat{\bf
n}^{}_0\rangle$, then one has [Eq. (\ref{chipm})] \begin{align}
|c^{}_+|^2=|\langle \hat{\bf n}|\hat{\bf
n}^{}_0\rangle|^2=\frac{1}{2}\bigl (1+\hat{\bf
n}^{}_0\cdot\hat{\bf n}\bigr ).\label{nzn}
\end{align}
Thus, the charge transmission through the filter will decrease
from $T^{}_{+}(\epsilon)$, when $\hat{\bf n}^{}_0=\hat{\bf n}$, to
zero, when $\hat{\bf n}^{}_0=-\hat{\bf n}$. The relative magnitude
of this transmission, $T(\epsilon)/T^{}_{+}(\epsilon)=|c^{}_+|^2$
is a linear combination of the components of $\hat{\bf n}^{}_0$,
which can thus be extracted by measurements at three pre-tuned
values of $\hat{\bf n}$.  This amounts to a {\it reading} of the
polarization of the incoming electrons, namely of the quantum
information stored in these mobile qubits. If the incoming
electrons are not fully polarized, the measured charge
transmission will yield information on the average over
$|c^{}_+|^2$. Specifically, for random polarizations one has
$T=T^{}_+/2$. An alternative way to test the polarization of the
outgoing spins is to send them through another filter, see Subsec.
IIIB.

\subsection{Single diamond with Rashba SOI}\label{SUBSEC:RSOI-only}

For specific types of interaction one needs explicit forms for the
unitary hopping matrices $U^{}_{uv}$. Consider a bond of length
$L$  from ${\bf r}^{}_u$ to ${\bf r}^{}_v$, with ${\bf r}_v-{\bf
r}_u\equiv L{\hat {\bf g}}_{uv}$. Placing the magnetic field ${\bf
H}$ along the $z$ direction and choosing the gauge ${\bf
A}=\frac{1}{2}{\bf H}\times {\bf r}$, we assign an AB phase
$\phi^{}_{uv}$ to $U^{}_{uv}$:
\begin{align}
\phi^{}_{uv}=\frac{\pi HL}{\Phi_0}[\hat{\bf g}_{uv}\times{\hat
z}]\cdot{\bf r}_u.\label{phiuv}
\end{align}

With the SOI, one has $U^{}_{uv}=\exp[i\phi^{}_{uv}+i{\bf
K}^{}_{uv}\cdot\sigmav]$, and ${\bf K}^{}_{uv}$ is given by Eq.
(\ref{oreg1}) with $\hat{\bf g}\rightarrow\hat{\bf g}^{}_{uv}$. To
demonstrate the power of our formalism, we start here by
considering only the Rashba SOI. Generalizing Ref.
~\onlinecite{hatano}, our diamond is a rhombus with an opening
angle of $2\beta$. Choosing the $x-$axis along the leads, the four
sites of the diamond are at ${\bf r}_0=(0,0,0)$, ${\bf
r}_b=(L\cos\beta,L\sin\beta,0)$, ${\bf
r}_c=(L\cos\beta,-L\sin\beta,0)$ and ${\bf r}_1=(2L\cos\beta,0,0)$
(Fig. 1).
 The hopping matrices for the four bonds
then become
\begin{align} U^{}_{0b}&=\exp (i\alpha\sigma^{}_1),\ \ \
U^{}_{b1}=\exp(-i\phi/2-i\alpha\sigma^{}_2),\nonumber\\
U^{}_{0c}&=\exp (-i\alpha\sigma^{}_2),\ \ \
U^{}_{c1}=\exp(i\phi/2+i\alpha\sigma^{}_1),\label{eq23}
\end{align}
where $\alpha=\alpha^{}_R=k^{}_{R}L\equiv \alpha^{}_1/\cos\beta$,
$\sigma^{}_1=\sin\beta\sigma^{}_x-\cos\beta\sigma^{}_y$,
$\sigma^{}_2=\sin\beta\sigma^{}_x+\cos\beta\sigma^{}_y$ and
$\phi/(2\pi)= H L^2 \sin(2\beta)/\Phi_0\equiv\phi^{}_1\tan\beta$
is the number of flux units through the diamond. In these
expressions we have introduced
\begin{align}
\alpha^{}_1=k^{}_RL^{}_0,\ \ \ \phi^{}_1=2 H L_0^2/\Phi^{}_0,
\label{a1f1} \end{align} where $2L^{}_0$ is the distance between
sites 0 and 1. These parameters do not depend on $\beta$ even when
the sites $b$ and $c$ are moved in order to vary $\beta$ (see
below). Substituting the matrices (\ref{eq23}) into Eq. (\ref{WW})
one obtains Eq. (\ref{WW1}), with
\begin{align}
&d=a^{}_+[c^2-s^2\cos(2\beta)],\ \ \ b^{}_x=0,\nonumber\\
&b^{}_y=-2ia^{}_+ c s \cos\beta,\ \ \
b^{}_z=ia^{}_-s^2\sin(2\beta),\label{db}
\end{align}
where  $c=\cos\alpha$, $s=\sin\alpha$ and
\begin{align}
a^{}_\pm&=\gamma^{}_b e^{-i\phi/2}\pm\gamma^{}_c e^{i\phi/2}.
\label{eq01}
\end{align}
Equation (\ref{WWA}) now reproduces Eq. (\ref{AB}), with the
identification $\cos\omega=1-2s^4\sin^2(2\beta)$ and
$\sin\omega=2s^2\sin(2\beta)\sqrt{1-s^4\sin^2(2\beta)}$. For
$\beta=\pi/4$ this value of $\omega$ was found in Ref.
~\onlinecite{hatano}. Also,
\begin{align}
\hat{\bf n}=s^{}_\phi\bigl (2c s
\cos\beta,0,c^2-s^2\cos(2\beta)\bigr
)/\sqrt{1-s^4\sin^2(2\beta)},\label{eq13}
\end{align}
where $s^{}_\phi={\rm sign}[\sin\phi]$. Below we present results
for $s^{}_\phi>0$. Furthermore, Eq. (\ref{Bprime}) gives
\begin{align}
\hat{\bf n}'=(-\hat{n}^{}_x,0,\hat{n}^{}_z).\label{eq14}
\end{align}
The condition (\ref{condi}) now implies that
\begin{align}
\cos(\phi/2)=\pm \sin^2\alpha\sin(2\beta).\label{cond}
\end{align}
 This equation
corresponds to a line in the $\phi$-$\alpha$ plane, which is shown
in Fig. \ref{222} (for three values of $\beta$). If one varies
both $\alpha$ and $\phi$ along such a line, then the outgoing
spins are fully polarized, and the transmission is given by Eq.
(\ref{Tp}). One should note that $\phi=2\pi \phi^{}_1\tan\beta$
and $\alpha=\alpha^{}_1/\cos\beta$, and these relations should be
taken into account when translating $\phi$ and $\alpha$ to the
magnetic and electric fields for different angles $\beta$.

\begin{figure}[h]
\includegraphics[width=6 cm]{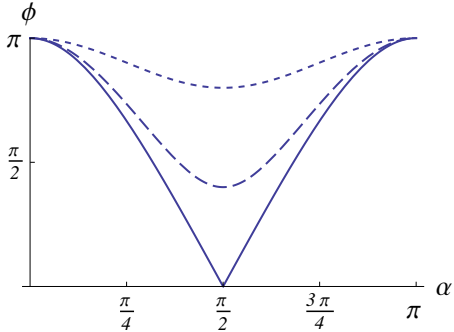}
 \caption{The relation between the AB flux $\phi$ and the Rashba SOI strength $\alpha$ at full filtering
 [Eq. (\ref{cond})].
 The full line and the dashed lines with decreasing sized dashes correspond to the rhombus's opening angle
 $\beta/\pi=.25,\ .15$ and $.05$.
 Results are the same
 under $(\pi/4-\beta)\leftrightarrow (\beta-\pi/4)$.}\label{222}
\end{figure}

When Eq. (\ref{cond}) is satisfied, the outgoing electrons are
polarized along $\hat{\bf n}'$. The variation of the components of
this polarization with $\alpha$, when one moves along the lines in
Fig. \ref{222}, is shown in Fig. \ref{3}, which also shows the
spin directions in the $xz-$plane for $\beta=\pi/4$. Changing the
sign of $\phi$ interchanges $|\hat{\bf n}\rangle$ and $|-\hat{\bf
n}\rangle$, and therefore changes the polarization associated with
the blocked spinor from $-\hat{\bf n}$ to $\hat{\bf n}$.

\begin{figure}[h]
\includegraphics[width=4.1 cm]{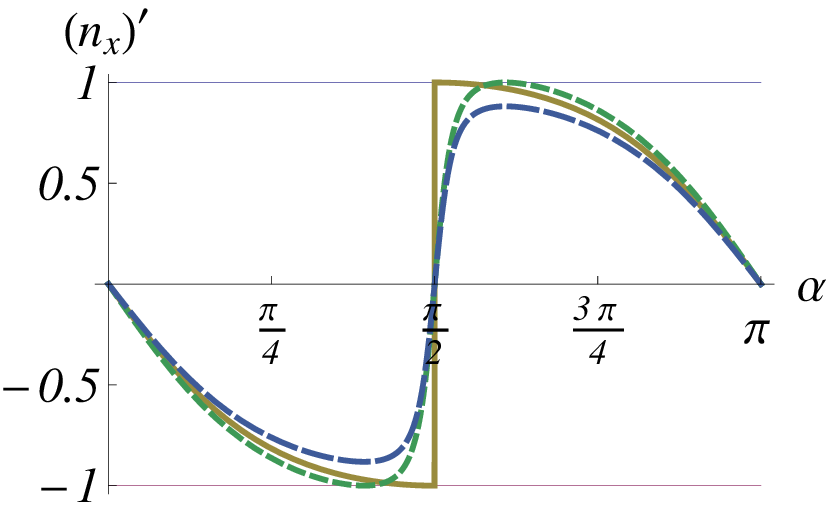}\ \ \includegraphics[width=4.1 cm]{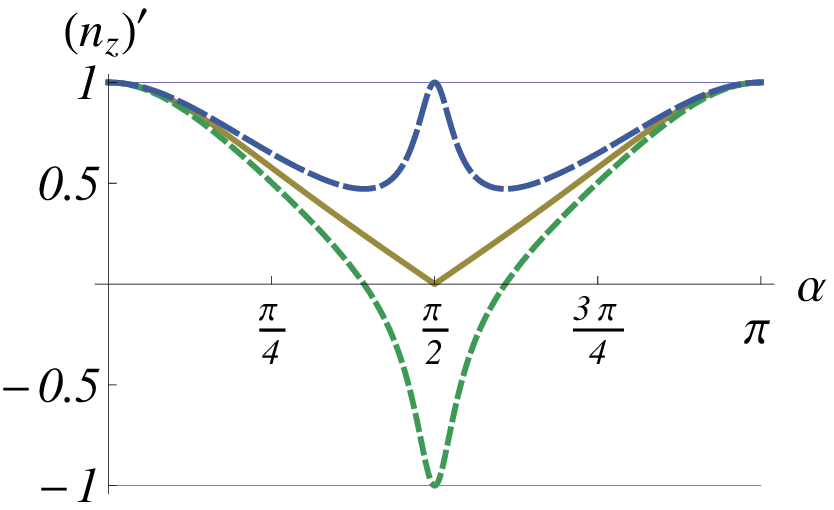}\\
\vspace{.3cm}
\includegraphics[width=6 cm]{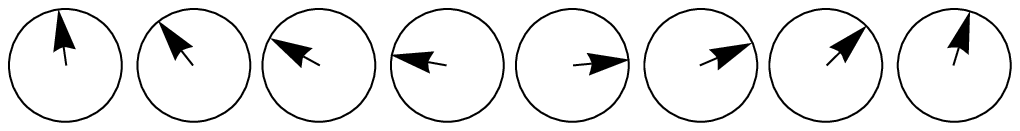}\\
 \caption{The outgoing spin components  for rhombus angles $\beta=.25\pi,\ .23\pi$ and $.27\pi$
 (full line, small dashes and large dashes, respectively), as a function of the Rashba SOI strength $\alpha$, when the AB flux is
 given by Eq. (\ref{cond}).
 Changing $\phi$ to $-\phi$
 switches the direction of the polarized spins. The
 lower panel shows the actual spin directions in the $xz-$plane for $\beta=\pi/4$, as  $\alpha$ increases from
 zero to $\pi$ (left to right).
 }\label{3}
\end{figure}

As Fig. \ref{3} shows, at small $\alpha$, namely small electric
field (and correspondingly large $\phi$, see Fig. \ref{222}) the
outgoing spin points along the $z-$axis, parallel to the magnetic
field. However, this spin ordering is not due to the Zeeman
interaction, which is small (and neglected here), but rather due
to the orbital effect of the AB flux. As $\alpha$ increases, the
spin rotates towards the negative $x-$direction. For
$\beta=\pi/4$, the spin reaches this direction as $\alpha
\rightarrow \pi/2$, and then flips abruptly to the opposite
direction. Upon further increase of $\alpha$, the outgoing spin
rotates back towards the positive $z-$direction. When $\beta \ne
\pi/4$, the outgoing spin also rotates towards the positive
$x-$direction, but now the results depend on $\beta$: when
$\beta<\pi/4$ ($>\pi/4$) the spin continuously rotates towards the
negative (positive) $z-$direction as $\alpha \rightarrow\pi/2$.
These dips (peaks) in $n'_z$ near $\alpha=\pi/2$ become sharper as
$\beta$ approaches $\pi/4$.

As mentioned in the previous subsection, there is no filtering at
exactly $\alpha=\pi/2$ and  $\beta=\pi/4$, namely at $\phi=0$.
However, the effect is most striking in the vicinity of
$\beta=\pi/4$, $\alpha=\pi/2$ and $\phi=0$. In fact, if one wishes
to flip the outgoing spins by a small change in the electric
field, which determines $\alpha$, then it would be best to use the
filter for a small finite flux $\phi$ and for $\beta=\pi/4$.
Changing $\alpha$ from $\pi/2-\phi\sqrt{2}/4$ to
$\pi/2+\phi\sqrt{2}/4$ will cause a jump in $n'_x$ from $-1$ to
$1$, i.e. a flip of the polarization from the negative to the
positive $x-$direction. Alternatively, two filters with
$\beta>\pi/4$ and $\beta<\pi/4$ would give opposite spin
components near $\alpha=\pi/2$ and the appropriate value of $\phi$
as given by Eq. (\ref{cond}).

Since $n^{}_y=0$, the procedure outlined after Eq. (\ref{nzn})
yields only the $x$ and $z$ components of the polarization of the
incoming electrons, $\hat{\bf n}^{}_0$. However, the third
component can always be deduced from $|\hat{\bf n}^{}_0|^2=1$. As
discussed below, this issue is overcome when one adds the
Dresselhous SOI. Alternatively, we note that the vector $\hat{\bf
n}$ is in the $xz-$plane only when the diamond is placed as in
Fig. 1, with the sites 0 and 1 on the $x-$axis. Placing these
sites along the $y-$axis will place $\hat{\bf n}$ in the
$yz-$plane. Thus, splitting the incoming beam, which contains many
electrons in identical spin states, into two beams, which go
through two diamonds placed along the two axes, will allow a
simultaneous determination of all the components of $\hat{\bf
n}^{}_0$.

So far, we applied Eq. (\ref{cond}) at fixed $\beta$, and obtained
full filtering by varying both $\phi$ and $\alpha$ (namely the
magnetic and electric fields) simultaneously. An alternative,
which may be more attractive under some circumstances, is to vary
$\beta$ by moving the dots $b$ and $c$ towards the $x-$axis. As
noted after Eq. (\ref{eq23}), such motion  also affects the area
of the diamond and the length of each edge. Fixing the magnetic
field fixes $\phi^{}_1$ [Eq. (\ref{a1f1})], and Eq. (\ref{cond})
becomes
\begin{align}
\alpha^{}_1=\pm\cos\beta\arccos\bigl
[1-2\cos(\phi^{}_1\tan\beta/2)/\sin(2\beta)\bigr ]/2.\label{condb}
\end{align}
The top panel in Fig. \ref{vbeta} shows this relation  for four
values of $\phi^{}_1$. All four lines show a smooth monotonic
variation of $\alpha^{}_1$ [i.e. the electric field responsible
for $k^{}_R$, Eq. (\ref{a1f1})] with $\beta$ (i.e. the electric
field responsible for moving the dots $b$ and $c$), over ranges
which become wider as $\phi^{}_1$ increases. The other panels in
Fig. \ref{vbeta} show the two components of the polarized spin
when Eq. (\ref{condb}) is obeyed. Varying $\beta$  rotates this
polarization. Setting the site energies so that the maximum of
$T^{}_+$ is at $\phi^{}_0=\phi^{}_1$, the transmission given in
Eq. (\ref{TTphi}) remains close to unity over a range of $\beta$
around $\pi/4$.

\begin{figure}[h]
\includegraphics[width=5 cm]{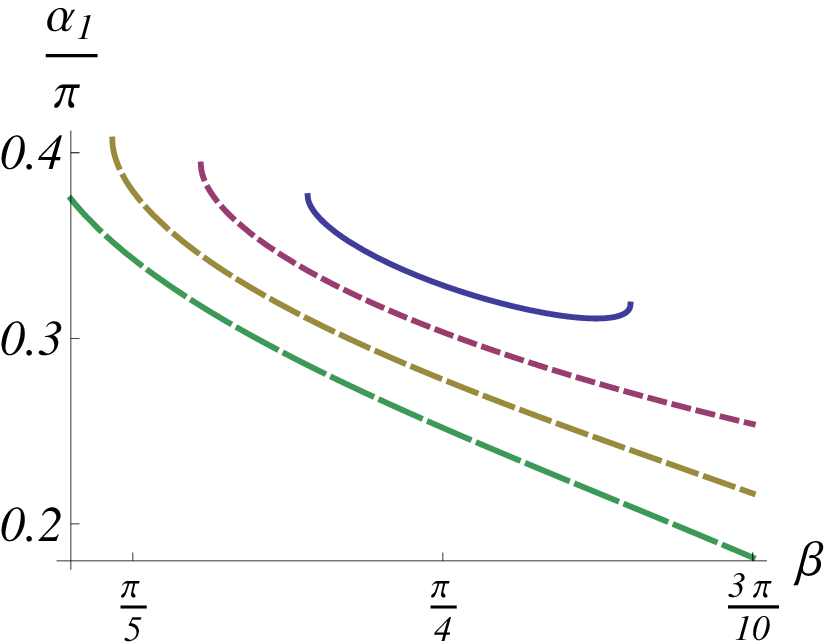}\\\vspace{.3cm}
\includegraphics[width=4 cm]{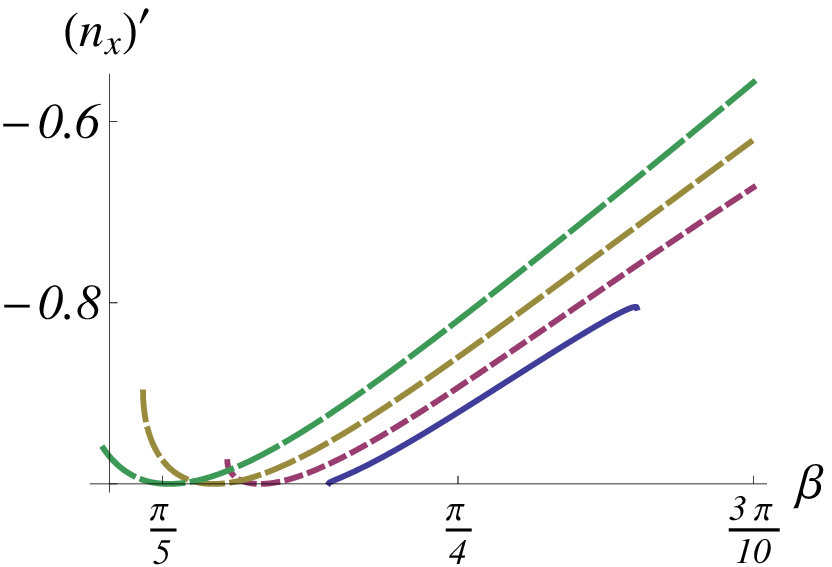}\ \ 
\includegraphics[width=4 cm]{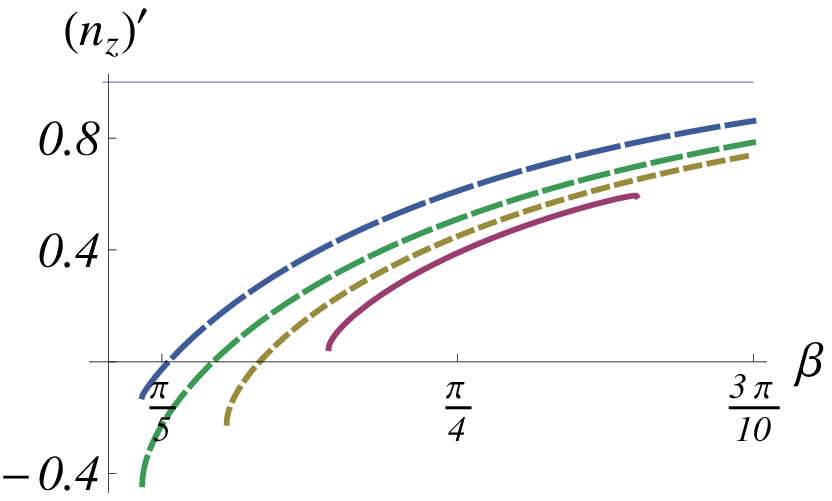}\\
 \caption{Top: The relation between the strength of the Rashba SOI $\alpha^{}_1=k^{}_RL_0$ and the
 diamond angle $\beta$  for full filtering, Eq. (\ref{condb}). Bottom: the two
 components of the outgoing polarized spins versus $\beta$. Full line and increasing  dashes
 correspond to $\phi^{}_1=.05,~.1,~.15$ and $.2$.}
\label{vbeta}
\end{figure}

\subsection{Single diamond with both Rashba and Dresselhaus SOI's}

 Since the Dresselhaus SOI [Eq. (\ref{Dress})] depends on the
directions of the crystal axes, one needs to introduce the angles
between the diamond bonds and these axes, as in Fig. {\ref{5}.
Thus,
$\mathbf{r}^{}_0=(0,0,0),\
\mathbf{r}^{}_1=2L\cos\beta(\cos\nu,\sin\nu,0),\ 
\mathbf{r}^{}_b=L(\cos(\nu+\beta),\sin(\nu+\beta),0),\ 
\mathbf{r}^{}_c=L(\cos(\nu-\beta),\sin(\nu-\beta),0)$.
We then use the vector ${\bf K}$ from Eq. (\ref{oreg1}) and the AB pahse from Eq. (\ref{phiuv}). Denoting
also
\begin{align}
&\zeta^2=\alpha_R^2+\alpha_D^2, \ \ \ \
\tan\theta=\alpha^{}_D/\alpha^{}_R,
\end{align}
 one recovers Eqs. (\ref{eq23}), with the
replacement of $\alpha$ by $\zeta$ and with the new spin
components 
$\sigma^{}_1\equiv \sin \xi^{}_1 \sigma^{}_x-\cos \xi^{}_2
\sigma_y,\ \sigma^{}_2\equiv \sin \xi^{}_4 \sigma^{}_x+\cos
\xi^{}_3\sigma^{}_y$, 
 where $\xi^{}_1\equiv \beta+\nu+\theta,$
$\xi^{}_2\equiv \beta+\nu-\theta,$ $\xi^{}_3\equiv
\beta-\nu+\theta,$ and $\xi^{}_4\equiv \beta-\nu-\theta$.

Note that $\sigma_1^2=F^2_1=1+\sin(2\nu+2\beta)\sin(2\theta)$,
$\sigma_2^2=F^2_2=1+\sin(2\nu-2\beta)\sin(2\theta)$, and therefore
$e^{i\zeta\sigma_n}=c^{}_n+is^{}_n\sigma^{}_n$, with
$c^{}_n\equiv\cos(\zeta F^{}_n),~s^{}_n\equiv\sin(\zeta
F^{}_n)/F^{}_n$. With these notations, one has
\begin{align}
e^{i\zeta\sigma^{}_1}e^{-i\zeta\sigma^{}_2}=\delta+i\tauv\cdot\sigmav,\
\ \
e^{-i\zeta\sigma^{}_2}e^{i\zeta\sigma^{}_1}=\delta+i\tauv'\cdot\sigmav,\end{align}
where
\begin{align}
&\delta=c^{}_1c^{}_2+s^{}_1s^{}_2(\sin\xi^{}_1\sin\xi^{}_4-\cos\xi^{}_2\cos\xi^{}_3),\nonumber\\
&\tau^{}_{x}=\tau'_{x}=s^{}_1c^{}_2\sin\xi^{}_1 -
c^{}_1s^{}_2\sin\xi^{}_4,\nonumber\\
&\tau^{}_{y}=\tau'_{y}=-s^{}_1c^{}_2\cos\xi^{}_2-c^{}_1s^{}_2\cos\xi^{}_3,\nonumber\\
&\tau^{}_{z}=-\tau'_{z}=s^{}_1s^{}_2(\sin\xi^{}_1\cos\xi^{}_3+\cos\xi^{}_2\sin\xi^{}_4),
\end{align}
and $\delta^2+|\tauv|^2=1$ from unitarity. It is now
straightforward to recover the matrix ${\bf W}$ [Eq. (\ref{WW1})],
with
\begin{align}
d=a_+\delta,\ \ \ b^{}_x=ia^{}_+\tau^{}_{x},\ \
b^{}_y=ia^{}_+\tau^{}_{y},\ \ \ b^{}_z=ia^{}_-\tau^{}_{z},
\end{align}
and with $a^{}_\pm$ as given in Eq. (\ref{eq01}). Again, Eq.
(\ref{WWA}) is used to recover Eq. (\ref{AB}), with the
identifications $\cos\omega=1-2\tau_{z}^2$ and \begin{align}
\hat{\bf n}=(-\tau^{}_{y},\tau^{}_{x},\delta)/\sqrt{1-\tau_{z}^2}.
\end{align}

The condition for full filtering, Eq. (\ref{condi}), now becomes
\begin{align}
\cos(\phi/2)=\pm\tau^{}_{z}=\pm
s^{}_1s^{}_2\sin(2\beta)\cos(2\theta).\label{eq27}
\end{align}
This is the main result of this subsection. One immediately notes
the following. (a) This condition reduces to Eq. (\ref{cond}) when
$\alpha^{}_D=0$. (b) When $\alpha^{}_R=0$, this condition also
reduces to Eq. (\ref{cond}), with $\alpha^{}_D$ replacing
$\alpha^{}_R$. This is not surprising, since the two types of SOI
are related via a unitary transformation. (c) When
$\alpha^{}_D=\pm \alpha^{}_R$ then $\cos(2\theta)=0$, and
therefore Eq. (\ref{eq27}) yields $\phi=\pi$, i.e. $\sin\phi=0$.
As discussed following Eq. (\ref{lam+}), there is no filtering in
this case, and all electrons are fully reflected.

When both $\alpha^{}_D$ and $\alpha^{}_R$ have non-trivial values
then all three components of $\hat{\bf n}$ are non-zero, and
therefore the procedure outlined after Eq. (\ref{nzn}) allows the
determination of {\it all} the three components of $\hat{\bf
n}^{}_0$ by measuring the charge transmission for three values of
$\hat{\bf n}$. As mentioned, usually $k^{}_D$ is fixed for a given
material and $k^{}_R$ can be varied experimentally by tuning the
electric field in the $z$ direction. At non-trivial values of
$k^{}_D$, the values of $\phi$ for full filtering, Eq.
(\ref{eq27}), are no longer periodic in $\alpha^{}_R$. An
exception occurs for
 $\nu=0$ and
$\beta=\pi/4$ (see Fig. \ref{5}), when (\ref{eq27}) reduces to
\begin{align}
\pm\cos(\phi/2)=\sin^2\alpha^{}_R-\sin^2\alpha^{}_D.\label{nu0}
\end{align}
Figure \ref{phiRD} shows this special periodic result for three
values of $\alpha^{}_D$ (full and larger dashes). Interestingly,
if $\alpha^{}_D=\pi/2$ then the curve for the pure Rashba SOI case
just shifts to the left by $\pi/2$, so that the interesting regime
moves to small electric and magnetic fields. However,  other
values of $\alpha^{}_D$ (e.g. $\pi/4$ in the figure) give a much
narrower range of $\phi$.

For all other values of $\nu$ and $\beta$ the flux $\phi$ for full
filtering is not periodic in $\alpha^{}_R$, and one must use the
full expression (\ref{eq27}). This expression  becomes simple for
$\nu=\beta=\pi/4$,
\begin{align}
\pm\cos(\phi/2)=\sin^2\zeta\cos(2\theta).\label{nupi}
\end{align}
Figure \ref{phiRD} also shows this function (smallest dashes) for
$\alpha^{}_D=\pi/4$. As $\alpha^{}_R$ increases, this expression
approaches the Rashba condition, Eq. (\ref{cond}), whereas Eq.
(\ref{nu0}) remains periodic in $\alpha^{}_R$. As seen in the
figure, working near $\alpha^{}_R=3\pi/2$ already brings us close
to the pure Rashba behavior. We trust that this value can be
achieved with reasonable electric fields. Presumably, one can
control the angle $\nu$ by rotating  the crystal  which forms the
filter.

\begin{figure}[h]
\includegraphics[width=6 cm]{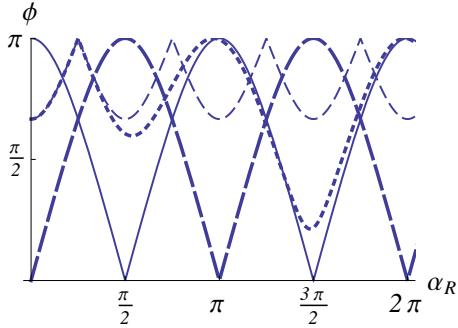}
 \caption{Same as Fig. \ref{222}, with $\beta=\pi/4$, $\nu=0$ (as defined in Fig. \ref{5})
 and with the Dresselhaus SOI strength $\alpha^{}_D=0,~,\pi/4$ and $\pi/2$
 (full line, medium  and long dashes respectively), Eq. (\ref{nu0}). The line with the smallest dashes
 shows the filtering condition for $\beta=\nu=\pi/4$ and
 $\alpha^{}_D=\pi/4$, Eq. (\ref{nupi}).}\label{phiRD}
\end{figure}

\section{Two diamonds}

\subsection{General formalism}

Consider now two diamonds in series, connected at site 1 (Fig.
\ref{2f}, top). Eliminating the side sites ($b,~c,~d$ and $e$), we
have
\begin{align}
&z^{}_0|\psi(0)\rangle={\bf
W}_A|\psi(1)\rangle-j|\psi(-1)\rangle,\nonumber\\
&z^{}_2|\psi(2)\rangle={\bf
W}_B^\dagger|\psi(1)\rangle-j|\psi(3)\rangle,\nonumber\\
&Z^{}_1|\psi(1)\rangle={\bf W}_A^\dagger|\psi(0)\rangle+{\bf
W}^{}_B|\psi(2)\rangle, \label{123}
\end{align}
where $z^{}_0=\epsilon-y^{}_0$ [Eq. (\ref{defz})],
$z^{}_2=\epsilon-\epsilon^{}_2-\gamma^{}_{2d2}-\gamma^{}_{2e2}$,
$Z^{}_1=\epsilon-\epsilon^{}_1-\gamma^{}_{1b1}-\gamma^{}_{1c1}-\gamma^{}_{1d1}-\gamma^{}_{1e1}$,
 ${\bf W}^{}_{A}$ represents Eq. (\ref{WW}) for diamond A and ${\bf W}^{}_B$ is defined similarly,
 with $d,e$ replacing $b,c$. Eliminating site $1$ then yields
\begin{align}
&(z^{}_0Z^{}_1{\bf 1}-{\bf W}^{}_A{\bf
W}^\dagger_A)|\psi(0)\rangle={\bf
W}^{}_A{\bf W}^{}_B|\psi(2)\rangle-Z^{}_1j|\psi(-1)\rangle,\nonumber\\
&(Z^{}_1z^{}_2{\bf 1}-{\bf W}^\dagger_B{\bf
W}^{}_B)|\psi(2)\rangle={\bf W}^\dagger_B{\bf
W}^\dagger_A|\psi(0)\rangle-Z^{}_1j|\psi(3)\rangle.
\end{align}

Using the analogs of Eq. (\ref{psi}) for a wave coming from the
left, and utilizing
 the identity \begin{align}
 {\bf W}^\dagger[C{\bf 1}+{\cal O}{\bf W}^\dagger]^{-1}\equiv[C{\bf 1}+{\bf
 W}^\dagger{\cal O}]^{-1}{\bf W}^\dagger,
 \end{align}
  where ${\cal O}$ and ${\bf W}$ are arbitrary $2 \times 2$ matrices and $C$ is a
 number, yields for the transmission amplitude from left to right,
\begin{widetext}\begin{align}
{\cal T}&=2ij\sin(ka){\bf W}^\dagger_B\bigl
[Z^{}_1X^{}_0X^{}_2{\bf 1}+X^{}_0{\bf W}^{}_B{\bf
W}^\dagger_B+X^{}_2{\bf W}^\dagger_A{\bf
W}^{}_A\bigr ]^{-1}{\bf W}^\dagger_A\nonumber\\
&=2ij\sin(ka){\bf
W}^\dagger_B\frac{Z^{}_1X^{}_0X^{}_2+X^{}_0A^{}_B+X^{}_2A^{}_A-(X^{}_0{\bf
B}^{}_B+X^{}_2{\bf
B}'^{}_A)\cdot\sigmav}{(Z^{}_1X^{}_0X^{}_2+X^{}_0A^{}_B+X^{}_2A^{}_A)^2-(X^{}_0{\bf
B}^{}_B+X^{}_2{\bf B}'^{}_A)^2}{\bf W}^\dagger_A,\label{T2d}
\end{align}
\end{widetext}
where the second step uses Eqs. (\ref{eq25}) and (\ref{eq25a}),
with the corresponding coefficients $A^{}_{A,B}$, ${\bf
B}^{}_{A,B}$ and ${\bf B}'^{}_{A,B}$. Here,
 $X^{}_0$ was defined in Eq.
(\ref{YX}), while similarly
$X^{}_2=\epsilon^{}_2+\gamma^{}_{2d2}+\gamma^{}_{2e2}+je^{-ika}$.

The factor ${\bf W}^\dagger_A$  on the RHS implies that if we
choose the parameters of $A$ to produce full polarization then all
the electrons entering from the left become fully polarized along
$\hat{\bf n}'_A$, and the other factors in ${\cal T}$ can be used
to tune the unique polarization of the outgoing electrons, and
perhaps the amplitude of the net transmission. Similarly, the
factor ${\bf W}^\dagger_B$ on the LHS means that if we tune $B$ to
give full polarization then the outgoing electrons will all be
polarized along $\hat{\bf n}'_B$, irrespective of the polarization
of the incoming electrons.

We next consider two diamonds with an additional bond between
them, see the lower panel in Fig. \ref{2f}. The tight-binding
equations are similar to the above, but now we also introduce a
SOI on the bond between the sites 1 and 2,
$\widetilde{U}^{}_{12}=J^{}_0U$, where $U$ is a unitary matrix to
be specified. Straightforward algebra, similar to that presented
above, yields the left to right transmission amplitude
\begin{widetext}
\begin{align}
{\cal T}=-2ij\sin(ka)J^{}_0{\bf W}_B^\dagger U^\dagger\bigl
[\Delta X^{}_0X^{}_3{\bf 1}+z^{}_1X^{}_0U{\bf W}^{}_B{\bf
W}^\dagger_BU^\dagger
+z^{}_2X^{}_3{\bf W}^\dagger_A{\bf
W}^{}_A+{\bf W}^\dagger_A{\bf W}^{}_AU{\bf W}^{}_B{\bf
W}^\dagger_BU^\dagger\bigr ]^{-1}{\bf W}^\dagger_A,\label{T2dU}
\end{align}
\end{widetext}
with $\Delta=z^{}_1z^{}_2-J_0^2$,
$z^{}_2=\epsilon-\epsilon^{}_2-\gamma^{}_{2d2}-\gamma^{}_{2e2}$,
$X^{}_0$ from Eq. (\ref{YX}) and
$X^{}_3=\epsilon^{}_3+\gamma^{}_{3d3}+\gamma^{}_{3e3}+je^{-ika}$.

Interestingly, one again has ${\bf W}^\dagger_A$ on the RHS and
${\bf W}^\dagger_B$ on the LHS. We expect this to be the case for
any structure with these two diamonds at the ends. However, the
present case differs from the previous one, since now ${\bf
W}^{}_B$ appears only in the combination $\widetilde{\bf
W}^{}_B=U{\bf W}^{}_B$. Therefore one can use the unitary rotation
of the spins $U$ to modify the states which enter the diamond $B$,
as discussed below.

\subsection{Ideal filtering}

If both $A$ and $B$ are tuned to be full polarizers then one has
\begin{align}
{\bf W}^\dagger_{A}=\sqrt{\lambda^{}_{A+}}|\hat{\bf
n}'_A\rangle\langle\hat{\bf n}^{}_A|,\ \ \ {\bf
W}^\dagger_{B}=\sqrt{\lambda^{}_{B+}}|\hat{\bf
n}'_B\rangle\langle\hat{\bf n}^{}_B|\label{chW}
\end{align}
and therefore Eq. (\ref{T2d}) reduces to
\begin{align}
{\cal T}=\frac{2ij\sin(ka)\langle\hat{\bf n}^{}_B|\hat{\bf
n}'^{}_A\rangle\sqrt{\lambda^{}_{A+}\lambda^{}_{B+}}}{Z^{}_1X^{}_0X^{}_2
+X^{}_0\lambda^{}_{B+}+X^{}_2\lambda^{}_{A+}}|\hat{\bf
n}'_B\rangle\langle \hat{\bf n}^{}_A|.\label{T2s}
\end{align}
More generally, we can choose only one of the diamonds to fully
polarize, and then we can tune the other one for optimization of
the transmission and/or for tuning the polarization through the
other diamond.

A particularly interesting possibility is to choose two identical
diamonds, with
$\beta^{}_A=\beta^{}_B,~\epsilon^{}_b=\epsilon^{}_c=\epsilon^{}_d=\epsilon^{}_d$
and $J^{}_{uv}\equiv J$. The only free parameters are now the two
$\phi$'s and the strengths of the SOI's. Consider the pure Rashba
case, and choose also  $\phi^{}_A=\phi^{}_B$ and
$\pi/2-\alpha^{}_A=\alpha^{}_B-\pi/2$, namely
$\sin(2\alpha^{}_A)=-\sin(2\alpha^{}_B)$. With these choices one
has $A^{}_A=A^{}_B$ and ${\bf B}^{}_A={\bf B}'^{}_B$ [see Eqs.
(\ref{eq13}) and (\ref{eq14})]. In this case, the transmission
amplitude is
\begin{align} {\cal
T}=\frac{2ij\sin(ka)\lambda^{}_+}{Z^{}_1X^{}_0X^{}_2+(X^{}_0+X^{}_2)\lambda^{}_+}
|\hat{\bf n}^{}_A\rangle\langle \hat{\bf n}^{}_A|.
\end{align}
This device has the advantage that it fully transmits electrons
with polarization along $\hat{\bf n}^{}_A$, and does not rotate
them as the single-diamond filter. Another important advantage
involves the transmission. When $Z^{}_1$ is very small (but
non-zero) and if also $\epsilon \simeq 0$ and
$\epsilon^{}_0+2\gamma^{}_b=\epsilon^{}_2+2\gamma^{}_d=0$ then the
transmission of this polarized spin is close to unity for almost
all $\phi$'s except for a narrow range near $\phi=0$ or $\pi$
(${\cal T}$ still vanishes when $\sin\phi=0$). Moving away from
the band center, the shape of the transmission $T(\epsilon)$
depends on $J$ and on $\epsilon^{}_b$. For large enough $J$, the
$\gamma$'s depend only weakly on $\epsilon$, and  the transmission
approaches the trivial value $\sin^2(ka)$ ($=1$ in the band
center), coming from the velocity of the electrons in the band.
Thus, this structure is an ideal polarizer for energies close to
the band center.

If we choose ${\bf B}^{}_A=-{\bf B}'^{}_B$ then all the electrons
will be blocked. For Rashba SOI, the latter condition implies the
relations
$\hat{n}^{}_{A,x}=\hat{n}^{}_{B,x},~\hat{n}^{}_{A,z}=-\hat{n}^{}_{B,z}$,
which can be realized if both
$\sin(2\alpha^{}_A)=-\sin(2\alpha^{}_B)$ and
$\phi^{}_A=-\phi^{}_B$ [Eqs. (\ref{AB}), (\ref{eq13})].

Finally, consider  the second double diamond device, lower Fig.
\ref{2f}. Since $U$ is unitary, the eigenvalues of $\widetilde{\bf
W}^{}_B\widetilde{\bf W}^\dagger_B$ are the same as for ${\bf
W}^{}_B{\bf W}^\dagger_B$, namely $\lambda^{}_{B\pm}$. However,
the eigenstates are different: $|\pm\hat{\bf
n}^{}_{BU}\rangle=U|\pm\hat{\bf n}^{}_{B}\rangle$. Using ${\bf
W}^\dagger_A$ from Eq. (\ref{chW}) and $\widetilde{\bf
W}^\dagger_B=\sqrt{\lambda^{}_{B+}}|\hat{\bf
n}'_{BU}\rangle\langle \hat{\bf n}^{}_{BU}|$, Eq, (\ref{T2dU})
becomes ${\cal T}=t|\hat{\bf n}'_{BU}\rangle\langle \hat{\bf
n}^{}_A|$, with the transmission amplitude
\begin{align}
t=\frac{-2ij\sin(ka)J^{}_0\langle\hat{\bf n}^{}_{BU}|\hat{\bf
n}'_{A}\rangle\sqrt{\lambda^{}_{A+}\lambda^{}_{B+}}}{\Delta
X^{}_0X^{}_3+z^{}_1X^{}_0\lambda^{}_{B+}+z^{}_2X^{}_3\lambda^{}_{A+}+\lambda^{}_{A+}\lambda^{}_{B+}}.
\end{align}
The choices
$\epsilon^{}_0=\epsilon^{}_1=2J^2/\epsilon^{}_b=2J^2/\epsilon^{}_c,
~\epsilon^{}_2=\epsilon^{}_3=2J^2/\epsilon^{}_d=2J^2/\epsilon^{}_e$,
$\hat{\bf n}^{}_{BU}=\hat{\bf n}'_A$ and
$J^{}_0=4J^4\sin(\phi^{}_{A0})\sin(\phi^{}_{B0})/(j\epsilon^{}_b\epsilon^{}_d)$
yield a flat maximum of $T=|t|^2$ at unity (similar to Fig.
\ref{Tpp}) for energies near the band center and fluxes near
$\phi^{}_{A0}$ and $\phi^{}_{B0}$. One can now tune the outgoing
polarization via ${\bf W}^{}_A,~{\bf W}^{}_B$ and/or $U$.
Specifically, one has maximal transmission if one  tunes the
outgoing polarization to be along $\hat{\bf n}^{}_A$ by requiring
that $U|\hat{\bf n}'^{}_B\rangle=|\hat{\bf
n}'_{BU}\rangle=|\hat{\bf n}^{}_A\rangle$. In the Rashba case, the
vectors $\hat{\bf n}^{}_{A,B}$ and $\hat{\bf n}'_{A,B}$ are both
in the $xz-$plane, and therefore
$U=\exp[i\alpha^{}_{12}\sigma^{}_y]$, where
$2\alpha^{}_{12}=\arcsin([\hat{\bf n}^{}_B\times\hat{\bf
n}'_A]^{}_y)$. This rotation can be generated by an electric field
in the $z-$direction, and its magnitude $\alpha^{}_{12}$ can also
be changed by changing the length of the bond $12$. Since the two diamonds fully polarize the electrons, and the intermediate bond can rotate their polarization, this device can perform as the Datta-Das SFET.

\section{Summary and Discussion}

We have demonstrated that single-  and  double-diamond devices,
made of materials with strong SOI's, can act as both a spin filter
and a spin analyzer. Our calculations includes the following
specific achievements:

$\bullet$ Full filtering through a general single-loop
interferometer requires a symmetry between the two branches
($\gamma^{}_b=\gamma^{}_c$) and a relation between the AB phase
and the SOI phase ($\phi=\omega+\pi$). For the diamonds in Fig. 1
and in Fig. \ref{5} these conditions and the direction of the
filtered polarization $\hat{\bf n}'$ can be independent of the
electron's energy.

$\bullet$ The site energies and the hopping strengths around the
diamond can be chosen so that the transmission $T$ of the polarized
electrons is close to unity over a wide range of energies and AB
flux (Fig. \ref{Tpp}).

$\bullet$ The charge transmission of polarized spins through the
interferometer measures the angle between the directions of this
incoming polarization $\hat{\bf n}^{}_0$ and that characterizing the full
transmission by the filter, Eq. (\ref{nzn}), rendering a reading
of the former polarization.

$\bullet$ For the Rashba SOI, one can work at small magnetic
fluxes, and generate a flipping of the transmitted polarization by
a small change in the electric field (Fig. \ref{3}). One can also tune
the transmitted polarization keeping the magnetic field fixed, and
varying the shape of the diamond (Fig. \ref{vbeta}).

$\bullet$ Adding the Dresselhaus SOI usually breaks the
periodicity in the Rashba SOI strength $k^{}_R$ of the filtering
criterion, and complicates the various expressions. However,
increasing $k^{}_R$ brings the various expressions back to the
pure Rashba ones.

$\bullet$ The two-diamond device can be tuned to be symmetric, so
that the polarization of the electrons exiting the two-diamond
device is equal to that of the incoming ones (from either side),
with a transmission close to unity.

$\bullet$ Adding a bond with a SOI between the two diamonds allows
tuning of the polarization of the spins. This adds much
flexibility in the choice of the two diamonds. Since each diamond
acts as a full filter, this double-diamond device achieves the
aims of the Datta-Das SFET without ferromagnetic leads.

Are there materials for which one can reach values of
$\alpha^{}_R$ or order $\pi/2$, as required here? A Shubnikov-de
Haas experiment\cite{Grunder} on an Al$_{0.25}$In$_{0.75}$As
barrier layer gave a value for the Rashba coefficient (in
different units) $\alpha=3\times 10^{-11}$eV/m. With the effective
mass $m^\ast=0.023m^{}_0$, this gives
$k^{}_R=m^\ast\alpha/\hbar^2=9\times 10^6{\rm m}^{-1}$. Weak
antilocalization measurements in a quaternary InGaAsP/InGaAs
heterointerface\cite{kohda} yielded $\alpha=10.4\times
10^{-12}$eVm. With an effective mass $m^\ast=0.0408m^{}_0$, this
gives $k^{}_R=5.55\times 10^6{\rm m}^{-1}$. Thus, $L=300$nm would
imply $\alpha^{}_R\sim 1.6-2.7$,
allowing for $\alpha^{}_R=\pi/2$. 

As in most of our references, we calculated only the {\it
transmission} from left to right (or from right to left). Indeed,
the results for the transmission describe the outcome of {\it
scattering experiments}, when one has a beam of electrons coming
in only from one side of the device. In many experiments, one
would like to measure the {\it conductance} between these two
sides, which involves the difference between a current coming from
the left and a current coming from the right, as originally
discussed by Landauer.\cite{L70} A generalization of this approach
for our case has the form\cite{cylinder}
\begin{align}
I^j=\int
\frac{d\epsilon}{2\pi}[f^{}_L(\epsilon)-f^{}_R(\epsilon)]{\rm
Tr}[{\cal T}{\cal T}^\dagger \sigma^{}_j],
\end{align}
where
$f_{L,R}(\epsilon)=1/[1+e^{(\epsilon-\mu^{}_{L,R})/k^{}_BT}]$ is
the Fermi distribution function in the left $L$ or right $R$
reservoir, $T$ is the  temperature, and $\mu^{}_{L,R}$ are the
chemical potentials on the electronic reservoirs connected to the
leads). Denoting $\sigma^{}_0=1$, $I^0$ gives the net charge
current in units of $e/h$. Denoting the Pauli matrices by
$\sigma^{}_j,~j=1,2,3$, the corresponding vector ${\bf I}\equiv
(I^1,I^2,I^3)$ gives the net spin current in the leads. In all of
our examples, the hermitian matrix ${\cal T}{\cal T}^\dagger $ can
be written as ${\cal T}{\cal T}^\dagger =T^{}_+|\hat{\bf
n}'\rangle\langle\hat{\bf n}'|+T^{}_-|-\hat{\bf n}'\rangle\langle
-\hat{\bf n}'|$, where $\hat{\bf n}'$ denotes the polarization of
the outgoing electrons [see Eq. (\ref{eigenT})]. Therefore, at
linear response and at zero temperature the charge conductance  is
given by $(e^2/h)(T^{}_++T^{}_-)$, and the spin conductance is
${\bf I}=(T^{}_+-T^{}_-)\hat{\bf n}'$. When $T^{}_-=0$, both
currents are associated with $T^{}_+$, and the current is {\it
fully polarized} even at linear response. When $\phi\pm\omega=\pi/2$ 
then $T^{}_+=T^{}_-=T^{}_0$, and the current is not polarized at all.
In this latter case the linear charge conductance is $2e^2T^{}_0/h$, 
which could reach the full quantum value of $2e^2/h$.

Finite temperatures or
bias voltages $eV=\mu^{}_L-\mu^{}_R$ require summing of
$T^{}_+(\epsilon)$ over energies. As noted, such sums do not
affect the condition for full filtering or the direction of
outgoing polarization. Furthermore, a flat energy dependence, as
in Fig. \ref{Tpp}, maintains a large conductance even after the
summation is carried over. This generalized Landauer formula still
gives only the currents between the two electron reservoirs, which
have {\it unpolarized} electrons. Therefore, the results are not
sensitive to the polarization of the incoming electrons, and a
device based on connecting two unpolarized reservoirs cannot
function as a spin analyzer. The situation changes for polarized
reservoirs, but this requires more research.

Our calculations were restricted to one-dimensional bonds between
the quantum dots. Real quantum wires may have a finite width.
Although one still has only a few relevant channels,\cite{kat}
their effect requires more analysis. In any case, the number of
channels can also be tuned by the gate voltages.

A major question, relevant to all filters, concerns the
experimental verification that the outgoing spins are indeed fully
polarized. One way to test this is to use the double-diamond
device, as discussed in Sec. III. Switching from full transmission
to no transmission by switching the sign of the magnetic field on
the second diamond will supply a proof that the electrons have
been polarized. An alternative way is to introduce a quantum dot
with a strong Coulomb interaction on or near the outgoing lead.
\cite{Iye,taruch} Starting with no occupation on this dot, and
then increasing the gate voltage on it to capture one electron
from the polarized flow, will block the current due to Pauli's
principle. Yet another method detects the polarized current in
quantum-point contacts via transverse electron
focusing.\cite{detect}

\vspace{.3cm}
 \acknowledgements We acknowledge discussions with Y.
Imry. AA and OEW acknowledge the hospitality of NTT and of the
ISSP, where this project started, and support from the ISF and
from the DIP. YT, AA and OEW also acknowledge support at NTT from
the Funding Program for World-Leading Innovative R and D on
Science and Technology (FIRST).

\end{document}